\documentclass[%
 reprint,
 amsmath,amssymb,
 aps,
floatfix,
]{revtex4-2}

\newcommand{\fullcirc}{\mbox{{\Large$\bullet$}}}

\newcommand{\fullsquare}{\mbox{$\blacksquare$}}

\newcommand{\dotted}{\protect\mbox{${\mathinner{\cdotp\cdotp\cdotp\cdotp\cdotp\cdotp}}$}}
\newcommand{\dashed}{\protect\mbox{-\ -\ -\ -}}

\newcommand{\full}{\protect\mbox{------}}

\usepackage{graphicx}
\usepackage{dcolumn}
\usepackage{bm}
\usepackage{color}

\usepackage{hyperref}
\hypersetup{
	breaklinks=true,   
}

\begin{document}

\preprint{APS/123-QED}

\title{Verification of Wave Turbulence Theory in the Kinetic Limit}

\author{Alexander Hrabski }
 \email{ahrabski@umich.edu}
\author{Yulin Pan}%
\affiliation{%
Department of Naval Architecture and Marine Engineering, University of Michigan, Ann Arbor, Michigan 48109, USA
}%

\date{\today}

\begin{abstract} 
Using the 1D Majda-McLaughlin-Tabak model as an example, we develop numerical experiments to study the validity of the Wave Kinetic Equation (WKE) at the kinetic limit (i.e., small nonlinearity and large domain). We show that the dynamics converges to the WKE prediction, in terms of the closure model and energy flux, when the kinetic limit is approached. When the kinetic limit is combined with a process of widening the inertial range, the theoretical Kolmogorov constant can be recovered numerically to a very high precision.
\end{abstract}

\maketitle


\section{Introduction}
Wave Turbulence (WT) describes the out-of-equilibrium statistical dynamics of multi-scale wave
systems. The theory of WT has been successfully applied to various physical contexts, including ocean surface waves \cite{hasselmann_non-linear_1962,zakharov_stability_1968}, internal gravity waves \cite{lvov_oceanic_2010,wu_energy_2023}, quantum turbulence \cite{nazarenko_wave_2006}, and gravitational waves in the early universe \cite{galtier_direct_2021}. For a given system, a statistical closure model can be developed that connects the high-order correlators of the wave field to pair correlators. When the closure model is taken in the kinetic limit, i.e., infinitesimal wave amplitude in an infinite domain, a wave kinetic equation (WKE) can be derived which describes the spectral evolution via a Boltzmann-like collision integral over wave-wave interactions. 

One of the most important features of the WKE is that it yields stationary power-law solutions with constant flux, known as Kolmogorov-Zakharov (KZ) spectra. For direct cascades, the general form of the KZ solution of wave action spectrum can be written as $n_k=CP^{\alpha}k^\gamma$, where $k$ is the wavenumber, $P$ is the energy flux, and $\alpha=1/2$ and $\alpha=1/3$ for systems with three and four-wave resonances, respectively. $C$ and $\gamma$ are constants that can be calculated as a part of this solution. Attempts to verify the KZ solution heavily focus on the scaling exponent $\gamma$ \cite{majda_one-dimensional_1997,zhang_numerical_2021,du_impact_2023,korotkevich_inverse_2023}, with only a handful of them targeting the Kolmogorov constant $C$. Among the latter, recent experimental validations \cite[e.g.,][]{deike_energy_2014,deike_role_2015} lead to larger discrepancies with theory (as much as a factor of $20$ difference), partly due to the challenge of precisely measuring $P$ from experimental data. The studies that are relatively successful consider well-controlled numerical simulations, including turbulence of capillary waves \cite{pan_direct_2014,pan_understanding_2017,pan_understanding_2017-1} and Bose-Einstein Condensates \cite{zhu_direct_2023}, which provides values of $C$ respectively about $40\%$ above and $10\%$ below the corresponding theoretical values. These two results signify that for a given power-law spectrum with theoretical slope $\gamma$, the energy flux computed from simulations are respectively about $0.5$ and $1.3$ ($\approx (1/1.40)^2, 1.1^3$) times of that from the KZ solution.

The scarce and limited success in verifying the Kolmogorov constant indicates an insufficient understanding of the validity of the WKE for stationary WT. This issue is more subtle than verification of WKE at an evolving state \cite[e.g.,][]{zhu_testing_2022} where spectral evolution serves as a natural measure of success. We are mainly interested in two fundamental questions in this work: (1) What are the major obstacles in obtaining theoretical value of $C$ in simulations of dynamical equations? In particular, what does it take to bring the numerical value of $C$ further closer to the theoretical value than in previous validations? (2) How is KZ spectrum (and more generally, the WT closure) realized in one-dimensional (1D) systems? We note that 1D WT is in a sense more difficult to describe than in higher dimensions \cite{deng_full_2021,zhu_testing_2022,zhu_direct_2023} due to far fewer interactions for any given spectral range. One example of a 1D system is the Majda-McLaughlin-Tabak model \cite{majda_one-dimensional_1997}, for which the theoretical value of $\gamma$ has been notoriously difficult to reproduce numerically for more than 20 years. This puzzle was resolved only recently in \cite{du_impact_2023}, where it was shown that the theoretical value of $\gamma$ can be recovered with a much wider inertial range than those in previous studies. The Kolmogorov constant for this 1D system, on the other hand, has never been numerically studied.

With the two above questions in mind, we perform a numerical study of the MMT model focusing on the Kolmogorov constant in the stationary state. One feature distinguishing our current study from all previous studies is that we numerically probe the kinetic limit, by weakening nonlinearity while making the domain large. We show that the theoretical value of $C$ can be recovered as a result of two limiting processes: (1) as we approach the kinetic limit, the dynamics of the MMT model converge to the WKE description, evaluated through the closure model and energy flux; (2) as we enlarge the inertial range, the value of $C$ computed from the collision integral converges to the theoretical value to very high precision. In component (1), we find that quasi-resonances can lead to difference between MMT simulation and WKE prediction when the former is taken outside the kinetic limit as in \cite{zhu_direct_2023}. In component (2), we find that an inertial range of at least $3.5$ decades is needed for the convergence of $C$.

\section{The MMT Model and KZ Solution}

The MMT model is a family of nonlinear dispersive equations for a complex field $\psi(x)=\psi\in\mathbb{C}$ that are widely used in the study of WT \cite{majda_one-dimensional_1997,zakharov_one-dimensional_2004,chibbaro_weak_2017,hrabski_effect_2020}. The MMT equation of interest to this work reads
\begin{equation}
    i\frac{\partial \psi}{\partial t}=|\partial_x|^{1/2}\psi+|\psi|^2\psi,
    \label{eqn:mmt}
\end{equation}
where the derivative operator $|\partial_x|^{1/2}$ produces a dispersion relation $\omega_k=|k|^{1/2}$, the same as surface gravity waves. We consider the MMT equation on a 1D periodic domain of length $L$, where we have $\psi (t)=\sum_{k\in\Lambda_L}\hat{\psi}_{k}(t)e^{ikx}$, with $\Lambda_{L}\equiv 2\pi\mathbb{Z}/L$.

The statistical description of \eqref{eqn:mmt} begins by defining the wave action spectrum $n_{k} = \frac{L}{2\pi}\langle|\hat{\psi}_{k}|^{2}\rangle$, where the $\langle\cdot\rangle$ denotes an ensemble average (or a time average for statistically stationary data). In deriving the WKE governing $n_k$, a statistical closure needs to be taken which connects 4th-order correlation of $\hat{\psi}_{k}$ to $n_k$, in the form of
\begin{equation}
    \begin{split}
        \text{Im}\langle\hat{\psi}_{1}\hat{\psi}_{2}\hat{\psi}_{3}^*\hat{\psi}_{k}^*\rangle_{\Omega}=& 4\pi n_1 n_2 n_3 n_k \\ &\times \left(\frac{1}{n_k}+\frac{1}{n_3}-\frac{1}{n_1}-\frac{1}{n_2}\right)f(\Omega)
    \end{split}
    \label{eqn:closure}
\end{equation}
where $\Omega$ denotes the frequency mismatch of the four wave modes $k_1$, $k_2$, $k_3$ and $k$. The closure \eqref{eqn:closure} can be derived in various ways, assuming quasi-Gaussian statistics \cite{hasselmann_non-linear_1962}, or more recently under the less restrictive assumption of a field with random phases and amplitudes \cite{choi_joint_2005,eyink_kinetic_2012,chibbaro_4-wave_2018}. Depending on different methods of derivation, $f(\Omega)$ takes different forms of a broadened delta function, namely a Lorentzian form \cite{zakharov_kolmogorov_1992} or sinc-like functions \cite{nazarenko_wave_2011,onorato_straightforward_2020}. At the kinetic limit, we have $f(\Omega)\rightarrow \delta(\Omega)$ and reach the WKE.
\begin{equation}
    \begin{split}
        \frac{\partial n_{k}}{\partial t}=\iiint4\pi n_1 n_2 n_3 n_k \left(\frac{1}{n_k}+\frac{1}{n_3}-\frac{1}{n_1}-\frac{1}{n_2}\right) \\
        \times\delta( k_1 + k_2 - k_3 - k)\delta(\Omega)dk_1 dk_2 dk_3,   
    \end{split}
    \label{eqn:wke}
\end{equation}

One can further seek stationary solutions to \eqref{eqn:wke} using the so-called Zakharov transformation \cite{zakharov_kolmogorov_1992}. Of interest here is the KZ solution associated with a finite energy flux from small to large k, taking the form of $n_k = CP^{1/3}k^{-1}$. Included in the supplemental material \cite{supp} is a full derivation of the KZ spectrum, including a correction to the previous work \cite{zakharov_one-dimensional_2004} leading to a new value of $C=0.2984$.

\section{Methods of Numerical Study}

\begin{figure}
    \centering
    \includegraphics[width=8cm]{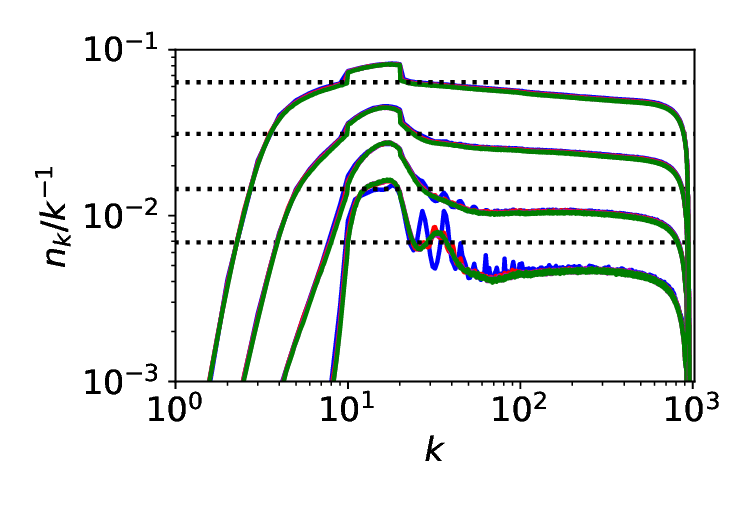}
    \caption{The directionally-averaged compensated wave action spectra $n_k/k^{-1}$ for each tested case. Colors denote $L=2\pi$ (blue), $L=4\pi$ (red), $L=8\pi$ (magenta), and $L=16\pi$ (green). The four distinct spectral levels denote the four values of $\varepsilon$. The KZ spectrum associated with $P(k_b=300)$ from the $L=16\pi$ case at each $\varepsilon$ is also plotted (\dotted).}
    \label{fig:spec}
\end{figure}

We simulate \eqref{eqn:mmt} via the pseudospectral method developed in \cite{majda_one-dimensional_1997}. To create a steady forward energy cascade, we add Gaussian forcing in the range of $10\le|k|\le20$ and a dissipation term $-i\nu_k\hat{\psi}_k$ to the RHS of (1), with $\nu$ given by
\begin{eqnarray}
     \nu_{k} = \left\{ \begin{array}{ll}
     3 k^{-4} & 0<|k|\le10 \\
    10 ^{-14} (k-900)^{8} & |k| \ge 900 \\
     0 & \text{otherwise}.
    \end{array} \right.
    \label{eqn:dissipation_largebox}
\end{eqnarray}
For each simulation, we start from a quiescent field and simulate until a stationary spectrum is reached, with the final nonlinearity level measured by $\epsilon=H_4/H_2$, where $H_2$ and $H_4$ are the linear and nonlinear components of total energy \cite{majda_one-dimensional_1997}, respectively.

To numerically approximate the kinetic limit, we choose four forcing strengths leading to four final nonlinearity levels, and for each forcing strength, we conduct simulations on domain of sizes $L\in[2\pi, 4\pi, 8\pi, 16\pi]$. As $L$ increases from $2\pi$ to $16\pi$, the resolution in $k$ space progressively doubles with $\Delta k$ decreasing from $1$ to $1/8$, with $k_{\text{max}}=1024$ kept for all simulations. In doing so extra care needs to be taken to scale the forcing and dissipation parameters with the domain size $L$, in order to keep quantities such as $\epsilon$ and total energy nearly constant across different domain sizes. Details on how this is done are included in the supplemental materials \cite{supp}.

With numerical data available, we are interested in studying the behavior of the closure model \eqref{eqn:closure}, especially in the context of a large number of quartets forming the energy flux. The basis for this analysis is an exact evaluation and decomposition of energy flux developed in \cite{hrabski_properties_2022}:
\begin{equation}
    \begin{split}
        P(k_b) =  \sum_{\Omega} P_\Omega(k_b),
    \end{split}
    \label{eqn:flux_decomp_def}
\end{equation}
\begin{equation}
        P_{\Omega} (k_b)=-\sum_{|k|<k_b} \omega_{k}\sum_{(k_{1}, k_{2}, k_{3})\in S_{\Omega , k}} 2\text{Im}\langle\hat{\psi}_{1}\hat{\psi}_{2}\hat{\psi}_{3}^*\hat{\psi}_{k}^*\rangle,   
    \label{eqn:pomega_def}
\end{equation}
where $k_b$ is the wave number through which the time-averaged flux $P$ is evaluated, $S_{\Omega , k} \equiv\ \{ (k_1, k_2, k_3) \big| \ k_1 + k_2 - k_3 - k = 0 , \  |\omega_{1}+\omega_{2}-\omega_{3}-\omega_{k}| = \Omega  \}$. Eq. \eqref{eqn:flux_decomp_def} describes the decomposition of energy flux into contributions from quartets with the wavenumber condition satisfied and with a mismatch of $\Omega$ in frequency condition. The formulation of $P_{\Omega}$ in \eqref{eqn:pomega_def} can be derived directly from \eqref{eqn:mmt} without any assumption \cite{hrabski_properties_2022}. With the closure model \eqref{eqn:closure} substituted in \eqref{eqn:pomega_def}, we have
\begin{equation}
    \begin{split}
        P_\Omega (k_b)=-\sum_{ \boldsymbol{k}\in \{ \boldsymbol{k}|k<k_b \} } \omega_{k}\sum_{(k_{1}, k_{2}, k_{3})\in S_{\Omega , k}} 4\pi n_1 n_2 n_3 n_k \\ 
        \times \left(\frac{1}{n_k}+\frac{1}{n_3}-\frac{1}{n_1}-\frac{1}{n_2}\right)f(\Omega),   
    \end{split}
    \label{eqn:pomega_closed}
\end{equation}
By computing the LHS of \eqref{eqn:pomega_closed} through \eqref{eqn:pomega_def} and the RHS via $n_k$ taken from our simulation data, we can directly compute the functional form of $f(\Omega)$. Comparison of the numerically-resolved $f(\Omega)$ with the analytical function then provides us a metric to evaluate the validity of the closure model. We note that this method evaluates the performance of \eqref{eqn:closure} over a large number of quartets, which is shown in \cite{hrabski_properties_2022} as the only meaningful way to study the closure model with a single simulation. In addition, since the RHS of \eqref{eqn:pomega_closed} is exactly the energy flux calculated from a discrete form of WKE with a broadened delta function, named quasi-resonant WKE (QRWKE) in \cite{pan_understanding_2017}, the closeness between numerical and analytical $f(\Omega)$ also indicates the accuracy of the WKE (or QRWKE) in reproducing the energy flux in dynamical simulations.

\section{Results}

\begin{figure}
    \centering
    \includegraphics[width=8cm]{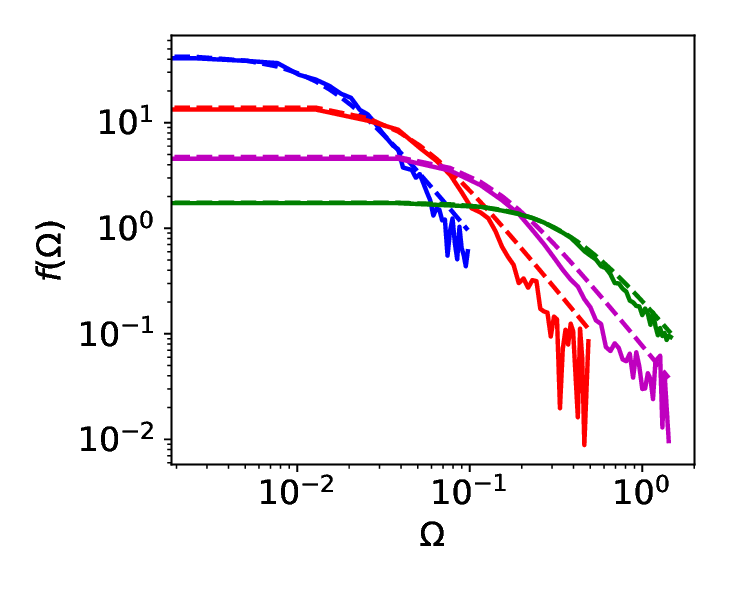}
    \caption{The measured closure function $f(\Omega)$ (\full) for each tested $L=16\pi$ case, denoted from highest to lowest $\varepsilon$ by green, magenta, red, and blue, respectively. A fitted Lorentzian closure $f(\Omega)$ for each case (\dashed).}
    \label{fig:closure}
\end{figure}

We begin by showing in figure \ref{fig:spec} the stationary spectra for all 16 simulations varying nonlinearity level $\epsilon\in(0.0066, 0.067)$ and domain size $L$. At high nonlinearity levels, we see good agreement in spectral form for all $L$, indicating that even the smallest domain size $L=2\pi$ is sufficient to capture the large-$L$ dynamics. At low nonlinearity (especially $\epsilon=0.0066$), however, we see that the spectrum varies substantially as $L$ changes. In particular, the secondary peaks occurring at small $L$ reflect finite-size effects. They disappear as $L$ increases, suggesting a transition from the discrete to the kinetic turbulence regime \cite{lvov_discrete_2010,zhang_numerical_2021}.

Also shown in figure \ref{fig:spec} are the KZ solutions, with $P$ computed though $k_b=300$ for $L=16\pi$ for each $\varepsilon$. While we see the spectral slope gets closer to the KZ value of $\gamma=-1$ as as $\varepsilon$ decreases, the inertial interval also shrinks, and perhaps even departs slightly from a true power-law. Additionally, the spectral level of the numerical solution does not get closer to the KZ solution as nonlinearity is decreased, indicating that the numerically-resolved Kolmogorov constant does not agree better with its theoretical value. We note that this is not in contradiction with the major theme of the paper. As discussed below, we need to consider two limiting processes to precisely reproduce the theoretical value of $C$: one taking the kinetic limit, and the other increasing the width of the inertial range.

Before discussing the two limiting processes in detail, we would like to briefly mention another set of important results that are made available due to our detailed analysis. As demonstrated in the supplemental material \cite{supp}, the energy flux follows a Gaussian distribution, whereas the energy dissipation rate follows a log-normal distribution. Both have a standard deviation that scales with $1/\sqrt{L}$. The log-normal distribution on the dissipation rate, first observed in wave turbulence here, is especially interesting as it has also been described in flow turbulence \cite{mandelbrot_intermittent_1974,pearson_log-normal_2018} with a sound theoretical explanation yet to be developed.

\begin{figure*} 
    \centering
    \includegraphics[width=8cm]{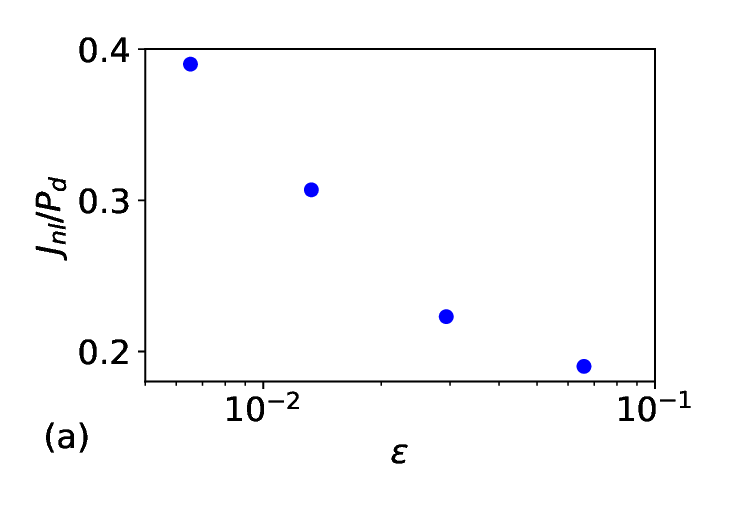}
    \includegraphics[width=8cm]{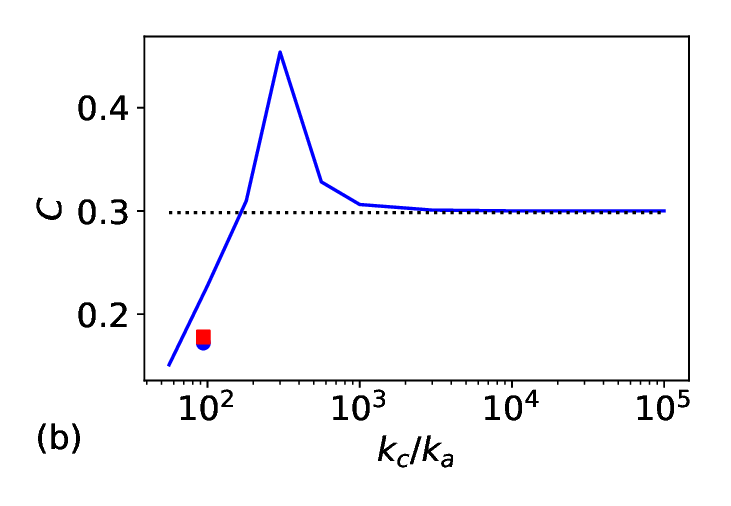}
    \caption{(a) The relative contribution of (non-local) interactions with the forcing range to the average small-scale dissipation rate for varying $\varepsilon$, plotted for the (representative) $L=16\pi$ case. (b) Value of $C$  for varying inertial interval length $k_c/k_a$. (\fullsquare) denotes $C$ computed via the kinetic flux for the simulated spectrum. (\fullcirc) denotes $C$ computed via dynamic flux for the simulated spectrum.}
    \label{fig:nonlocal}
\end{figure*}

\subsection{Limiting Process 1: Kinetic Limit}
In this section, we will show that as the kinetic limit is taken, the measured energy flux indeed converges to the prediction of the WKE. Following methods introduced in \S III, we plot $f(\Omega)$ for different nonlinearity levels at $L=16\pi$, together with the analytical Lorentzian form $a/\pi(a^2+\Omega^2)$ \cite{zakharov_kolmogorov_1992} with values of $a$ that best fit the data. We see that the analytical form fits the data remarkably well, and that as nonlinearity is decreased with sufficient domain size, the function $f(\Omega)$ approaches a true delta function. These results indicate that the long-time dynamics indeed follows the WT closure and converges to WKE description as the kinetic limit is taken. Due to the broadening of $f(\Omega)$, these results also suggest that dynamics away from the kinetic limit may be better represented by the QRWKE. Since the Lorentzian form of the delta functions contains long tails, it is expected that a large number of quasi-resonances are active in dynamic simulations away from the kinetic limit. These quasi-resonances are therefore key factors resulting in difference in energy flux from dynamical simulations and WKE calculations as seen in \cite{zhu_direct_2023}.

\subsection{Limiting process 2: Wide Inertial Range}

The convergence of dynamics to the WKE prediction at the kinetic limit does not guarantee that the simulated spectrum converges to the KZ solution, as we have demonstrated. As a result, the Kolmogorov constants evaluated for the simulated spectrum using the dynamic flux \eqref{eqn:pomega_def} and kinetic flux \eqref{eqn:pomega_closed} with $f(\Omega)\rightarrow\delta(\Omega)$ are very close to each other but away from the theoretical value (see symbols in figure \ref{fig:nonlocal}b). To understand this situation, recall that, given a valid WKE, the realization of the KZ solution requires the dominance of local interactions, which in turn requires a sufficiently wide inertial range. The width of inertial range seems to be especially important for 1D models such as MMT, as evidenced in \cite{du_impact_2023} on the sensitivity of $\gamma$ to the width. It is our objective to show next that our numerical solution is indeed contaminated by non-local interactions, and that a much wider inertial range is needed to precisely recover the theoretical value of $C$.

We first quantify the influence of forcing peak (as the source of non-local interactions) to the high-wavenumber portion of spectrum. Considering the stationary state, we explicitly decompose the average small-scale dissipation rate via $P_d = J_{nl} + J_{l}$, where $J$ is the small-scale energy evolution due to nonlinear interactions with subscripts $nl$ and $l$ denoting contributions from non-local and local interactions. An interaction is considered non-local if at least one wave number resides in forcing range $(10\le|k|\le20)$. Refining methods developed in \cite{simonis_time_2023} (see details in \cite{supp}), we explicitly compute $J_{nl}$ and plot the ratio $J_{nl}/P_d$ in figure \ref{fig:nonlocal}a. At low nonlinearities, about 40\% of the inter-scale energy flux is due to non-local interactions, clearly violating the assumptions of the KZ spectrum. Interestingly, the non-local contribution decreases as nonlinearity is increased, explaining the longer power-law range at higher nonlinearity seen in figure \ref{fig:spec}.

We finally demonstrate precise convergence to the theoretical value of $C$ as the width of inertial range is made larger. This is achieved by evaluating kinetic flux $P_{WKE}$ for an idealized spectrum $n_k = k^{-1}$ with cutoffs at both low and high wavenumbers $k_a$ and $k_c$. Here $P_{WKE}\equiv\int_{k_a}^{k_b}\omega_k \partial n_k /\partial t$ with $\partial n_k /\partial t$ computed from the collision integral of the WKE \eqref{eqn:wke}, which is exactly RHS of \eqref{eqn:pomega_closed} with $f(\Omega)\rightarrow\delta(\Omega)$. Specifically, we keep $k_a=10$ and progressively move $k_c$ to higher wavenumber to represent the widening of the inertial range, with $P_{WKE}$ evaluated at $k_b=560$. Figure \ref{fig:nonlocal}b plots the value of $C$ as a function of $k_c/k_a$. At $k_c/k_a \approx 100$, some discrepancy is seen between $C$ evaluated on the simulated and idealized spectra, due the differences in form between the two spectra. With the increase of $k_c/k_a$, we see that value of $C$ converges to the theoretical value with very high precision. We find that an inertial range of about 3.5 decade is needed to obtain converged and accurate result for $C$. This is much longer than what is needed to recover the theoretical $\gamma$ in MMT model \cite{du_impact_2023} and what is needed for $C$ in high-dimensional models \cite{zhu_direct_2023}. We note that for the MMT model we consider, $\gamma=-1$ is far from the boundaries of the locality window $-7/4<\gamma<1/2$ \cite{zakharov_one-dimensional_2004}. The long inertial range needed to recover the KZ solution is therefore more likely a general feature of the 1D wave turbulence.

\section{Conclusion}
We present a detailed numerical study on the MMT model, focusing on the realization of the Kolmogorov constant and the WT closure. We show that the Kolmogorov constant can be precisely recovered following two limiting processes: the first as kinetic limit is taken where the dynamics converges to the WKE prediction in terms of the WT closure model realized through energy flux; the second as the inertial-range spectrum is made sufficiently wide so that enough interactions over the inertial range are captured.

\begin{acknowledgments}
We thank Zaher Hani, Sergey Nazarenko, and Peter Miller for their thoughtful comments and suggestions. This work was supported by a grant from the Simons Foundation (Award ID \#651459, YP). This material is also based upon work supported by the National Science Foundation Graduate Research Fellowship under Grant No. DGE 1841052 (AH). Any opinions, findings, and conclusions or recommendations expressed in this material are those of the authors and do not necessarily reflect the views of the National Science Foundation. This work additionally used Bridges-2 at Pittsburgh Supercomputing Center through allocation phy220053p from the Advanced Cyberinfrastructure Coordination Ecosystem: Services \& Support (ACCESS) program, which is supported by National Science Foundation grants \#2138259, \#2138286, \#2138307, \#2137603, and \#2138296.
\end{acknowledgments}


\bibliography{HrabskiPanKinetic2023}

\end{document}


\newcommand{\fullcirc}{\mbox{{\Large$\bullet$}}}
\newcommand{\opensquare}{\mbox{$\square$}}
\newcommand{\opentriangle}{\mbox{$\vartriangle$}}
\newcommand{\opentriangledown}{\mbox{$\triangledown$}}
\newcommand{\opendiamond}{\mbox{$\lozenge$}}
\newcommand{\fullsquare}{\mbox{$\blacksquare$}}
\newcommand{\fulldiamond}{\mbox{$\blacklozenge$}}
\newcommand{\fullstar}{\mbox{$\bigstar$}}
\newcommand{\fulltriangle}{\mbox{$\blacktriangle$}}
\let\fulltri=\fulltriangle
\newcommand{\fulltriline}{\mbox{\textbf{---}$\blacktriangle$\textbf{---}}}
\newcommand{\fulltriangledown}{\mbox{$\blacktriangledown$}}
\let\fulltridown=\fulltriangledown
\newcommand{\opencirc}{\mbox{\Large$\circ$}}
\newcommand{\opensqr}{\mbox{$\square$}}
\newcommand{\opentri}{\mbox{$\triangle$}}
\newcommand{\dotted}{\protect\mbox{${\mathinner{\cdotp\cdotp\cdotp\cdotp\cdotp\cdotp}}$}}
\newcommand{\dashed}{\protect\mbox{-\ -\ -\ -}}
\newcommand{\broken}{\protect\mbox{-- -- --}}
\newcommand{\longbroken}{\protect\mbox{--- --- ---}}
\newcommand{\chain}{\protect\mbox{--- $\cdot$ ---}}
\newcommand{\dashddot}{\protect\mbox{--- $\cdot$ $\cdot$ ---}}
\newcommand{\full}{\protect\mbox{------}}

\title{Supplemental Material for the Paper \\ Verification of Wave Turbulence in the Kinetic Limit}

\author{Alexander Hrabski}
\author{Yulin Pan}
\affiliation{
Department of Naval Architecture and Marine Engineering, University of Michigan, Ann Arbor, Michigan 48109, USA
}

\maketitle

\section{Derivation of the Kolmogorov-Zakharov (KZ) Spectrum}

The derivation of the KZ spectrum begins with the Wave Kinetic Equation (WKE), provided again for convenience:
\begin{equation}
    \begin{split}
        \frac{\partial n_{k}}{\partial t}=\int\limits_{-\infty}^{\infty}\int\limits_{-\infty}^{\infty}\int\limits_{-\infty}^{\infty}4\pi n_1 n_2 n_3 n_k \left(\frac{1}{n_k}+\frac{1}{n_3}-\frac{1}{n_1}-\frac{1}{n_2}\right) \\
        \times\delta(k_1 + k_2 - k_3 - k)\delta(\omega_1 + \omega_2 - \omega_3 - \omega_k)dk_1 dk_2 dk_3.  
    \end{split}
    \label{eqn:wke_supp}
\end{equation}

While it isn't in general necessary \cite{nazarenko_wave_2011}, we adopt the popular assumption of an isotropic spectrum. In one-dimensional systems, this takes the form of $n_{k}=n_{-k}$. Each of $k_1$, $k_2$, and $k_3$ can take a positive or negative sign in the delta function of \eqref{eqn:wke_supp}, producing $8$ possibilities. However, not every combination of these signs produces a non-trivial resonance. A trivial resonance is one that satisfies the resonance conditions by having $k_1 = k_3$ and $k_2 = k$, or $k_2 = k_3$ and $k_1 = k$. For these cases, the integrand of \eqref{eqn:wke_supp} takes a zero value, so they may be ignored. After these trivial cases are removed, we are left with 
\begin{equation}
    \begin{split}
        \frac{\partial n_{k}}{\partial t} =  4\pi \int\limits_{0}^{\infty}\int\limits_{0}^{\infty}\int\limits_{0}^{\infty}n_{1}n_{2}n_{3}n_{k}\left(\frac{1}{n_{k}}+\frac{1}{n_{3}}-\frac{1}{n_{1}}-\frac{1}{n_{2}}\right)\delta(\omega_{1}+\omega_{2}-\omega_{3}-\omega_{k})\\
        \times\left(\delta(k_1 + k_2 + k_3 - k) +\delta(k_1 - k_2 - k_3 - k)+\delta(-k_1 + k_2 - k_3 - k)\right. \\
        +\left.\delta(-k_1 - k_2 + k_3 - k)\right)dk_1 dk_2 dk_3.
    \end{split}
    \label{eqn:mmt_wke_AI_supp}
\end{equation}
It will later become important to integrate over the resonant manifold (described by the $\delta$-functions), which is much easier to interpret as quadratic functions in $\omega=k^{1/2}$. Therefore, we next rewrite the above equation as an integral over $\omega$, and replace (on the LHS) $n_k$, the spectral density in $k$, with $\mathcal{N}_{\omega}$, the spectral density in $\omega$. This leads to
\begin{equation}
    \begin{split}
        \frac{\partial \mathcal{N}_{\omega}}{\partial t} =  128\pi \int\limits_{0}^{\infty}\int\limits_{0}^{\infty}\int\limits_{0}^{\infty}(\omega_1 \omega_2 \omega_3 \omega)n_{1}n_{2}n_{3}n_{\omega}\left(\frac{1}{n_{\omega}}+\frac{1}{n_{3}}-\frac{1}{n_{1}}-\frac{1}{n_{2}}\right)\delta(\omega_{1}+\omega_{2}-\omega_{3}-\omega)\\
        \times\left(\delta(\omega^2_1 + \omega^2_2 + \omega^2_3 - \omega^2) +\delta(\omega^2_1 - \omega^2_2 - \omega^2_3 - \omega^2)+\delta(-\omega^2_1 + \omega^2_2 - \omega^2_3 - \omega^2)\right. \\
        +\left.\delta(-\omega^2_1 - \omega^2_2 + \omega^2_3 - \omega^2)\right)d\omega_1 d\omega_2 d\omega_3 ,
    \end{split}
    \label{eqn:mmtw_wke_AI_supp}
\end{equation}
where $n_{\omega} = n(k(\omega))$, and we have used the facts that $dk = 2\omega d\omega$ and $\mathcal{N}(\omega)d\omega = n(k) dk + n(-k) dk = 2n(k)dk$. We note that the factor of $2$ on the spectral element relation is necessary for a consistent and correct flux definition.

We now assume that $n_\omega = A\omega^\gamma$, where $\gamma$ refers to an arbitrary exponent that will later be used to determine the KZ exponents. Substituting this into \eqref{eqn:mmtw_wke_AI_supp}, we are left with
\begin{equation}
    \begin{split}
        \frac{\partial \mathcal{N}_{\omega}}{\partial t} =  128\pi A^3 \int\limits_{0}^{\infty}\int\limits_{0}^{\infty}\int\limits_{0}^{\infty}\omega^{\gamma+1}_{1}\omega^{\gamma+1}_{2}\omega^{\gamma+1}_{3}\omega^{\gamma+1}\left(\omega^{-\gamma}+\omega^{-\gamma}_{3}-\omega^{-\gamma}_{1}-\omega^{-\gamma}_{2}\right)\delta(\omega_{1}+\omega_{2}-\omega_{3}-\omega)\\
        \times\left(\delta(\omega^2_1 + \omega^2_2 + \omega^2_3 - \omega^2) +\delta(\omega^2_1 - \omega^2_2 - \omega^2_3 - \omega^2)+\delta(-\omega^2_1 + \omega^2_2 - \omega^2_3 - \omega^2)\right. \\
        +\left.\delta(-\omega^2_1 - \omega^2_2 + \omega^2_3 - \omega^2)\right)d\omega_1 d\omega_2 d\omega_3 ,
    \end{split}
    \label{eqn:ZakTransform_SS_supp}
\end{equation}

Next we employ the Zakharov transformations \cite{zakharov_kolmogorov_1992,nazarenko_wave_2011}, which are a set of conformal transformations one applies to the integrand that result in the reduction of the sum of delta functions to a single delta function. This new structure of the integrand will allow us to (a) explicitly see the zeros of the equation and (b) explicitly compute the Kolmogorov constant $C$. See \cite{majda_one-dimensional_1997} for an intuitive, geometric description of how these transformations achieve these effects assuming only a self-similar spectrum. We distribute the sum of delta functions to expand the integrand into 4 terms, and we handle each separately. The first term,
\begin{equation}
    \begin{split}
        \frac{\partial \mathcal{N}_{\omega}}{\partial t}^{(1)} =  128\pi A^3 \int\limits_{0}^{\infty}\int\limits_{0}^{\infty}\int\limits_{0}^{\infty}\omega^{\gamma+1}_{1}\omega^{\gamma+1}_{2}\omega^{\gamma+1}_{3}\omega^{\gamma+1}\left(\omega^{-\gamma}+\omega^{-\gamma}_{3}-\omega^{-\gamma}_{1}-\omega^{-\gamma}_{2}\right) \\ 
        \times\delta(\omega_{1}+\omega_{2}-\omega_{3}-\omega)\delta(\omega^2_1 + \omega^2_2 + \omega^2_3 - \omega^2)d\omega_1 d\omega_2 d\omega_3 ,
    \end{split}
    \label{eqn:ZakTransform_T1_supp}
\end{equation}
is the form onto which we will map the other terms. We have used a superscript $(1)$ to denote we are referring to the first term. Now, as an example, we manipulate the second term in full,
\begin{equation}
    \begin{split}
        \frac{\partial \mathcal{N}_{\omega}}{\partial t} ^{(2)} =  128\pi A^3 \int\limits_{0}^{\infty}\int\limits_{0}^{\infty}\int\limits_{0}^{\infty}\omega^{\gamma+1}_{1}\omega^{\gamma+1}_{2}\omega^{\gamma+1}_{3}\omega^{\gamma+1}\left(\omega^{-\gamma}+\omega^{-\gamma}_{3}-\omega^{-\gamma}_{1}-\omega^{-\gamma}_{2}\right) \\
        \times\delta(\omega_{1}+\omega_{2}-\omega_{3}-\omega)\delta(\omega^2_1 - \omega^2_2 - \omega^2_3 - \omega^2)d\omega_1 d\omega_2 d\omega_3 ,
    \end{split}
    \label{eqn:ZakTransform_T2_supp}
\end{equation}
to which we apply the following transformations: $\omega_1 = \omega^2 / \omega_1 '$, $\omega_2 = \omega \omega_2 ' / \omega_1 '$, and $\omega_3 =  \omega \omega_2 ' / \omega_1 '$. Under these transformations, $d\omega_1 d\omega_2 d\omega_3 = \left(\frac{\omega}{\omega_1 '}\right)^4 d\omega_1 'd\omega_2 'd\omega_3 '$, and \eqref{eqn:ZakTransform_T2_supp} becomes, after some reduction,
\begin{equation}
    \begin{split}
        \frac{\partial \mathcal{N}_{\omega}}{\partial t}^{(2)} =  128\pi A^3  \int\limits_{0}^{\infty}\int\limits_{0}^{\infty}\int\limits_{0}^{\infty} \left(\frac{\omega
        }{\omega_1 '}\right)^{3\gamma + 5}\omega^{\prime \gamma+1}_{1} \omega^{\prime\gamma+1}_{2}\omega^{\prime\gamma+1}_{3}\omega^{\prime\gamma+1}\left(\omega^{\prime-\gamma}_{1}+\omega^{\prime-\gamma}_{3}-\omega^{\prime-\gamma}-\omega^{\prime-\gamma}_{2}\right) \\
        \times\delta(\omega'+\omega_{2}'-\omega_{3}'-\omega_{1}')\delta(\omega^{\prime2}_1 + \omega^{\prime2}_2 + \omega^{\prime2}_3 - \omega^{\prime2})d\omega_1 'd\omega_2 'd\omega_3 '.
    \end{split}
    \label{eqn:ZakTransform_T2_simple_supp}
\end{equation}
We note that the identity $\int\delta(a x)dx = \int \delta(x)/|a| dx$ is used to simply the above expression. The integrand of \eqref{eqn:ZakTransform_T2_simple_supp} \emph{almost} reflects \eqref{eqn:ZakTransform_T1_supp} with an additional factor of $(\omega/\omega_1 ')^{3\gamma +5}$. If one carefully looks at the signed terms in the equation, however, it becomes apparent that certain indices have become switched as a result of our transformation. The choice of indices is arbitrary, so we renumber them according to $123\rightarrow132$. This leaves us with
\begin{equation}
    \begin{split}
        \frac{\partial \mathcal{N}_{\omega}}{\partial t}^{(2)} =  -128\pi A^3  \int\limits_{0}^{\infty}\int\limits_{0}^{\infty}\int\limits_{0}^{\infty} \left(\frac{\omega
        }{\omega_1 '}\right)^{3\gamma + 5}\omega^{\prime \gamma+1}_{1} \omega^{\prime\gamma+1}_{2}\omega^{\prime\gamma+1}_{3}\omega^{\prime\gamma+1}\left(\omega^{\prime-\gamma}+\omega^{\prime-\gamma}_{3}-\omega^{\prime-\gamma}_{1}-\omega^{\prime-\gamma}_{2}\right) \\
        \times\delta(\omega'_{1}+\omega_{2}'-\omega_{3}'-\omega)\delta(\omega^{\prime2}_1 + \omega^{\prime2}_2 + \omega^{\prime2}_3 - \omega^{\prime2})d\omega_1 'd\omega_2 'd\omega_3 '.
    \end{split}
    \label{eqn:ZakTransform_T2_simpler_supp}
\end{equation}
In this form, the symmetry with \eqref{eqn:ZakTransform_T1_supp} is obvious. We perform the remaining Zakharov transformations (see \cite{zakharov_kolmogorov_1992} for the forms of the other transformations), sum the four terms, and drop the primes from our notation. This results in
\begin{equation}
    \begin{split}
        \frac{\partial \mathcal{N}_{\omega}}{\partial t} =  128\pi A^3 \int\limits_{0}^{\infty}\int\limits_{0}^{\infty}\int\limits_{0}^{\infty}\omega^{\gamma+1}_{1}\omega^{\gamma+1}_{2}\omega^{\gamma+1}_{3}\omega^{\gamma+1}\left(\omega^{-\gamma}+\omega^{-\gamma}_{3}-\omega^{-\gamma}_{1}-\omega^{-\gamma}_{2}\right) \\ 
        \times\left(1+\left(\frac{\omega_3}{\omega}\right)^{y}-\left(\frac{\omega_1}{\omega}\right)^{y}-\left(\frac{\omega_2}{\omega}\right)^{y}\right)\delta(\omega_{1}+\omega_{2}-\omega_{3}-\omega)\delta(\omega^2_1 + \omega^2_2 + \omega^2_3 - \omega^2)d\omega_1 d\omega_2 d\omega_3,
    \end{split}
    \label{eqn:ZakTransformed_supp}
\end{equation}
with $y\equiv -3\gamma - 5$. Careful consideration of the resonance conditions reveals that \eqref{eqn:ZakTransformed_supp} is an integral over the intersection of the plane $\omega_{1}+\omega_{2}-\omega_{3}-\omega = 0$ and the sphere $\omega^2_1 + \omega^2_2 + \omega^2_3 = \omega^2$ for any given $\omega$. Thus, we do not need to consider integrating over any $\omega_i > \omega$. This enables a reparameterization in terms of some $\xi_i  = \omega_i / \omega \in [0,1]$, so that \eqref{eqn:ZakTransformed_supp} becomes
\begin{equation}
    \begin{split}
        \frac{\partial \mathcal{N}_{\omega}}{\partial t} = 128\pi A^3 \omega^{-y-1} I(y) = 128\pi A^3 \omega^{-y-1} \int\limits_{0}^{1}\int\limits_{0}^{1}\int\limits_{0}^{1}\left(\xi_{1}\xi_{2}\xi_{3}\right)^{\gamma+1}\left(1+\xi^{-\gamma}_{3}-\xi^{-\gamma}_{1}-\xi^{-\gamma}_{2}\right) \\ 
        \times(1+\xi_{3}^{y}-\xi_{1}^{y}-\xi_{2}^{y})\delta(\xi_{1}+\xi_{2}-\xi_{3}-1)\delta(\xi^2_1 + \xi^2_2 + \xi^2_3 - 1) d\xi_1 d\xi_2 d\xi_3 .
    \end{split}
    \label{eqn:ZakTransform_final_supp}
\end{equation}
This is the form of the WKE that allows for the solution of $\gamma$ corresponding to stationary solutions. 

One can see by the second product in the integrand that if $\gamma = 0$, the RHS of \eqref{eqn:ZakTransform_final_supp} is identically $0$. This corresponds to equipartition of wave action, i.e., $n_\omega$ is constant. Also, if $\gamma = -1$, then the second product is identically $0$ whenever the resonance condition is satisfied, also leading to a $0$ of the collision integral. This solution $n_\omega = A/\omega$ corresponds to the Rayleigh-Jeans spectrum. Neither of these solutions correspond to wave turbulence, but rather are equilibrium solutions. The KZ solutions are given by $y = 0$ and $y=1$, which produce $0$'s of the collision integral by the same arguments as the equilibrium solutions. Setting $y=0$, one obtains $n_\omega = A\omega^{-5/3}$, and for $y=1$, one obtains $n_{\omega} = A\omega^{-2}$. To determine which of these out-of-equilibrium spectra correspond to the forward cascade of energy and which corresponds to the inverse cascade of wave action, one may use the ordering of the exponents $\gamma$ relative to the equilibrium spectra \cite{nazarenko_wave_2011}, or Zakharov's method via evaluating \eqref{eqn:ZakTransform_final_supp} \cite{zakharov_one-dimensional_2004}, while in both cases being careful to ensure the flux directions are not non-physical via comparison to the equilibrium spectra \cite{zakharov_kolmogorov_1992,nazarenko_wave_2011,zakharov_one-dimensional_2004}. In \cite{zakharov_one-dimensional_2004}, it is demonstrated that the forward and inverse cascade for our system have physical cascade directions and that $y=1$ (with $\gamma = -2$) corresponds to the forward cascade. 

Next, we derive the Kolmogorov constant $C$. The energy flux through frequency $\omega$ is defined by a control volume argument in spectral space as
\begin{equation}
    P(\omega) \equiv -\int\limits_{0}^{\omega} \omega' \frac{\partial \mathcal{N}}{\partial t}(\omega') d\omega' = -128\pi A^3 \frac{\omega^{1-y}}{1-y} I(y),
    \label{eqn:flux_def_for_C+theta_deri}
\end{equation}
where an negative sign is introduced to ensure that a positive flux corresponds to a cascade of energy from large to small scales (long to short time scales via the dispersion relation).  For the forward cascade with $y=1$, the computation of $P(\omega)$ involves the limit of an indeterminate quantity, which can be obtained via L'Hospital's rule to be
\begin{equation}
    P(\omega) = 128\pi A^3 \lim_{y\rightarrow1}\frac{d I(y)}{d y}.
    \label{eqn:flux_def_for_C+theta_deri_lhospital}
\end{equation}
The limit of the desired derivative, $S$, is given by
\begin{equation}
    \begin{split}
        S = \lim_{y\rightarrow1}\frac{d I(y)}{d y} = \int\limits_{0}^{1}\int\limits_{0}^{1}\int\limits_{0}^{1}\left(\xi_{1}\xi_{2}\xi_{3}\right)^{-1}\left(1+\xi^{2}_{3}-\xi^{2}_{1}-\xi^{2}_{2}\right) \\ 
        \times(\xi_{3}\ln{\xi_3}-\xi_{1}\ln{\xi_1}-\xi_{2}\ln{\xi_2})\delta(\xi_{1}+\xi_{2}-\xi_{3}-1)\delta(\xi^2_1 + \xi^2_2 + \xi^2_3 - 1) d\xi_1 d\xi_2 d\xi_3 ,
    \end{split}
    \label{eqn:dIdy}
\end{equation}
where we have used the fact that $\lim_{y\rightarrow1} \frac{d x^y}{dy} = x \ln{x}$. In the next section, we develop a precise method to numerically evaluate \eqref{eqn:dIdy} to find $S = 0.09353$. We now compute the relationship between $A$ and $P(\omega)$, revealing both the Kolmogorov constant $C$ via
\begin{equation}
        A = C P^{1/3} = \left(128\pi S\right)^{-1/3}P^{1/3},
    \label{eqn:kolmogorov_const_calc}
\end{equation}
resulting in $C = 0.2984$. Given that $C$ is positive, we can now be sure that the cascade direction is correct. Thus, the KZ spectrum associated with the forward cascade process in our MMT equation is given by $n_\omega = 0.2984P^{1/3}\omega^{-2}$, or, via the linear dispersion relation, $n_k = 0.2984P^{1/3}k^{-1}$. We note that this value of $C$ is different from the result of Zakharov et al. \cite{zakharov_one-dimensional_2004}, due to their missing factor of $2$ in the spectral element relation $\mathcal{N}(\omega)d\omega = 2n(k)dk$.

\section{Numerical Integration over the Resonant Manifold}

\subsection{Evaluation of $S$}
We are interested in integrating \eqref{eqn:dIdy}. For simplicity, we will refer the non-delta part of the integrand by  $f(\xi_1, \xi_2, \xi_3$) so that
\begin{equation}
    f(\xi_1 , \xi_2, \xi_3) \equiv \left(\xi_{1}\xi_{2}\xi_{3}\right)^{-1}\left(1+\xi^{2}_{3}-\xi^{2}_{1}-\xi^{2}_{2}\right)(\xi_{3}\ln{\xi_3}-\xi_{1}\ln{\xi_1}-\xi_{2}\ln{\xi_2}).
    \label{eqn:fdef_appendix}
\end{equation}
This leaves
\begin{equation}
    S = \int\limits_{0}^{1}\int\limits_{0}^{1}\int\limits_{0}^{1} f(\xi_1 , \xi_2, \xi_3)\delta(\xi_{1}+\xi_{2}-\xi_{3}-1)\delta(\xi^2_1 + \xi^2_2 + \xi^2_3 - 1) d\xi_1 d\xi_2 d\xi_3.
    \label{eqn:Sdef_appendix}
\end{equation}
The first of these delta functions is linear in $\xi_1$. Making use of the property
\begin{equation}
    \int\limits_{0}^{1} g(x) \delta(x-a) dx = g(a) \ \ \text{for} \ \ 0 \leq a \leq 1,
    \label{eqn:delta_prop}
\end{equation}
we can integrate over $\xi_3$ to obtain
\begin{equation}
    S = \iint\limits_{\Delta(\xi_1 , \xi_2)} f(\xi_1, \xi_2, \xi_1 + \xi_2 - 1)\delta\left(\xi_1^2 + \xi^2_2 + (\xi_1 + \xi_2 - 1)^2- 1\right) d\xi_1 d\xi_2.
    \label{eqn:S1_appendix}
\end{equation}
Integrating over the region $\Delta(\xi_1 , \xi_2) \equiv \{0<\xi_1 <1 , 0<\xi_2 <1 , 1<\xi_1 + \xi_2 <2 \}$ simply ensures $0<\xi_3 <1$. We would like to now apply \eqref{eqn:delta_prop} again, however we require a transformation so that the argument $\xi^2_1 + \xi^2_2 + \xi^2_3 - 1$ is of the required form. To do this, we transform the inner integral to one with respect to $du$, where $u = \xi_1^2 + \xi_2^2 + (\xi_1+\xi_2 - 1)^2 - 1$ and $du = 2(2\xi_1 + \xi_2 -1)d\xi_1$. This leaves
\begin{equation}
    S = \iint\limits_{\Delta(u,\xi_2)} \frac{f(\xi_1(u), \xi_2, \xi_1 (u) + \xi_2 - 1)}{2(2\xi_1 (u) + \xi_2 -1)}\delta\left(u\right) du d\xi_2.
    \label{eqn:S2_appendix}
\end{equation}
Now, we may apply \eqref{eqn:delta_prop}, being careful to include only the part of $u(\xi_1 , \xi_2)=0$ that lies in $\Delta(\xi_1 ,\xi_2)$. After some manipulation of our definition of $u$  we find that, of the two branches for which $u=0$, the one with
\begin{equation}
    \xi_1 = \frac{1 - \xi_2 + \sqrt{(1-\xi_2)(3\xi_2 + 1)}}{2}, \ \ 0<\xi_2<1
    \label{eqn:xi1ofu_appendix}
\end{equation}
is in the region $\Delta(\xi_1 ,\xi_2)$. After applying \eqref{eqn:delta_prop},
\begin{equation}
    S = \int\limits_{0}^{1} \frac{f(\xi_1(\xi_2), \xi_2, \xi_1 (\xi_2) + \xi_2 - 1)}{2\sqrt{(1-\xi_2)(3\xi_2 + 1)}} d\xi_2.
    \label{eqn:S3_appendix}
\end{equation}
This form of $S$ is suitable for numerical integration, where the integrand as $\xi_2 \rightarrow 1$ (from below) can be shown to approach zero via L'Hospital's rule.

\subsection{Evaluation of the WKE Collision Integral}

Just as in the evaluation of $\frac{dI(y)}{dy}$, this explicit algebraic approach works to evaluate \eqref{eqn:mmtw_wke_AI_supp}. To start, we rewrite \eqref{eqn:mmtw_wke_AI_supp} in terms of the function $f(\omega_1,\omega_2,\omega_3,\omega) = \omega_1 \omega_2 \omega_3 \omega n_{1}n_{2}n_{3}n_{\omega}\left(\frac{1}{n_{\omega}}+\frac{1}{n_{3}}-\frac{1}{n_{1}}-\frac{1}{n_{2}}\right)$:
\begin{equation}
    \begin{split}
        \frac{\partial \mathcal{N}_{\omega}}{\partial t} =  128\pi \int\limits_{0}^{\infty}\int\limits_{0}^{\infty}\int\limits_{0}^{\infty}f(\omega_1,\omega_2,\omega_3,\omega)\\
        \times\left(\delta(\omega^2_1 + \omega^2_2 + \omega^2_3 - \omega^2) +\delta(\omega^2_1 - \omega^2_2 - \omega^2_3 - \omega^2)+\delta(-\omega^2_1 + \omega^2_2 - \omega^2_3 - \omega^2)\right. \\
        +\left.\delta(-\omega^2_1 - \omega^2_2 + \omega^2_3 - \omega^2)\right)d\omega_1 d\omega_2 d\omega_3,
    \label{eqn:mmtw_wke_AI_f_supp}
    \end{split}
\end{equation}
For each $\delta$-function with quadratic arguments in $\omega$, we perform the procedure outlined in the previous subsection. After a good deal of simplification, this results in the following expression for $\frac{\partial \mathcal{N}_\omega}{\partial t}$:
\begin{eqnarray}
    \frac{\partial \mathcal{N}_\omega}{\partial t} =& \frac{\partial \mathcal{N}_\omega^{(1a)}}{\partial t} + \frac{\partial \mathcal{N}_\omega^{(1b)}}{\partial t} + \frac{\partial \mathcal{N}_\omega^{(2)}}{\partial t} + \frac{\partial \mathcal{N}_\omega^{(3a)}}{\partial t} + \frac{\partial \mathcal{N}_\omega^{(3b)}}{\partial t}, \text{ \ where} \\
    \frac{\partial \mathcal{N}_\omega^{(1a)}}{\partial t} =&  \int\limits_{0}^{\sqrt{4/3}\omega} \frac{f(\omega_1^{(1a)}(u),\omega_2^{(1a)}(u),\omega_3^{(1a)}(u),\omega)}{2\sqrt{4\omega^2-3u^2}} du \\
    \frac{\partial \mathcal{N}_\omega^{(1b)}}{\partial t} =&  \int\limits_{\omega}^{\sqrt{4/3}\omega} \frac{f(\omega_1^{(1b)}(u),\omega_2^{(1b)}(u),\omega_3^{(1b)}(u),\omega)}{2\sqrt{4\omega^2-3u^2}} du \\
    \frac{\partial \mathcal{N}_\omega^{(2)}}{\partial t} =&  2\int\limits_{0}^{\infty} \frac{f(\omega_1^{(2)}(u),\omega_2^{(2)}(u),\omega_3^{(2)}(u),\omega)}{2\sqrt{u^2+4\omega^2}} du \\
    \frac{\partial \mathcal{N}_\omega^{(3a)}}{\partial t} =&  \int\limits_{0}^{\infty} \frac{f(\omega_1^{(3a)}(u),\omega_2^{(3a)}(u),\omega_3^{(3a)}(u),\omega)}{2\sqrt{u^2+4\omega^2}} du \\
    \frac{\partial \mathcal{N}_\omega^{(3b)}}{\partial t} =&  \int\limits_{0}^{\infty} \frac{f(\omega_1^{(3b)}(u),\omega_2^{(3b)}(u),\omega_3^{(3b)}(u),\omega)}{2\sqrt{u^2+4\omega^2}} du.
\end{eqnarray}
The first delta function on the RHS of \eqref{eqn:mmtw_wke_AI_f_supp} corresponds to $\partial \mathcal{N}_\omega^{(1)} / \partial t$, the next two delta functions (due to symmetry) produce identical contributions and are combine in $\partial \mathcal{N}_\omega^{(2)} / \partial t$, and the last delta function corresponds to $\partial \mathcal{N}_\omega^{(3)} / \partial t$. The separate $a$ and $b$ contributions result from multiple branches of the resonant manifold parameterized by $u$. The corresponding functions of $\omega_i^{j}(u)$ are given by:
\begin{eqnarray}
    (1a) & \left\{ \begin{array}{l}
    \omega_2(u) = \frac{1}{3} \left(\omega + \sqrt{4\omega^2-3u^2}\right) \\
    \omega_1(u) = \frac{1}{2} \left(\omega - \omega_2(u) + \sqrt{(\omega-\omega_2(u))(3\omega_2(u)+\omega)}\right) \\
    \omega_3(u) = \omega_1(u) + \omega_2(u) - \omega
    \end{array} \right. \\
    (1b) & \left\{ \begin{array}{l}
    \omega_2(u) = \frac{1}{3} \left(\omega - \sqrt{4\omega^2-3u^2}\right) \\
    \omega_1(u) = \frac{1}{2} \left(\omega - \omega_2(u) + \sqrt{(\omega-\omega_2(u))(3\omega_2(u)+\omega)}\right) \\
    \omega_3(u) = \omega_1(u) + \omega_2(u) - \omega
    \end{array} \right. \\
    (2) & \left\{ \begin{array}{l}
    \omega_2(u) = -\omega + \sqrt{u^2+4\omega^2} \\
    \omega_1(u) = \frac{1}{2} \left(\omega - \omega_2 + \sqrt{(3\omega+\omega_2(u))(\omega_2(u)-\omega)}\right) \\
    \omega_3(u) = \omega_1(u) + \omega_2(u) - \omega
    \end{array} \right. \\
    (3a) & \left\{ \begin{array}{l}
    \omega_3(u) = \omega + \sqrt{u^2+4\omega^2} \\
    \omega_2(u) = \frac{1}{2} \left(\omega + \omega_3(u) + \sqrt{(\omega_3(u)+\omega)(\omega_3(u)-3\omega)}\right) \\
    \omega_1(u) = \omega_3(u) + \omega - \omega_2(u)
    \end{array} \right. \\
    (3b) & \left\{ \begin{array}{l}
    \omega_3(u) = \omega + \sqrt{u^2+4\omega^2} \\
    \omega_2(u) = \frac{1}{2} \left(\omega + \omega_3(u) - \sqrt{(\omega_3(u)+\omega)(\omega_3(u)-3\omega)}\right) \\
    \omega_1(u) = \omega_3(u) + \omega - \omega_2(u)
    \end{array} \right. \\
\end{eqnarray}
This is a form of the collision integral that is suitable for numerical integration. When computing $P_{WKE}$ directly from the collision integral in the main text, $n_k \equiv n(k(\omega))$ is evaluated explicitly from the idealized, truncated KZ spectrum given by
\begin{equation}
    n_k = \left\{ \begin{array}{l}
    k^{-1}, \ k_a \le k \le k_c  \\
    0, \ \text{otherwise}
    \end{array} \right. \\
\end{equation}
When the simulated spectrum is instead used in this evaluation (for the two points in figure 3b of the main text), we use piecewise-linear interpolation of the simulated spectrum to evaluate $n_k$ for $k\notin\Lambda_L$. In both of these cases, $n_k=0$ after some large value of $k$ with a corresponding $u$, meaning that the upper integral bounds of $\partial \mathcal{N}_\omega^{(3a)} / \partial t$ and $\partial \mathcal{N}_\omega^{(3b)} / \partial t$ in our numerical evaluations can be taken to be finite. All integrals are evaluated via adaptive quadrature. 

\section{Selection of Numerical Parameters}
For a fair comparison between domains of different $L$, we would like to keep key quantities like the length-averaged Hamiltonian density $H$, $\varepsilon$, and $P$ approximately constant as $L$ increases, while allowing other quantities to vary. Choosing the parameters of each numerical simulation to achieve this goal turns out to be a non-trivial task. Given the definition of our Fourier series
\begin{equation}
    \hat{\psi}_{k} = \frac{1}{L}\int\limits_{0}^{L} \psi_{\boldsymbol{x}}e^{-ik\cdot\boldsymbol{x}}d\boldsymbol{x},
    \label{eqn:ft}
\end{equation}
the spectral amplitude $\hat{\psi}_{k}$ is normalized such that the length-averaged total action does not depend on $L$, as can be seen via Parseval's theorem \cite{nazarenko_wave_2011},
\begin{equation}
    \frac{1}{L}\int\limits_{0}^{L} |\psi_{\boldsymbol{x}}|^2 d\boldsymbol{x} = \sum_{k \in \Lambda_L^2} |\hat{\psi}_{k}|^2.
    \label{eqn:parseval}
\end{equation}
The key parameters that control the energy content of our simulations are forcing and dissipation. Thus, we attempt to keep the action injection rate, controlled by the standard deviation of the Gaussian forcing $\sigma_F$, and the dissipation rate, controlled by $\nu_{k}$, the same across domains of different $L$ at each $\varepsilon$ of interest. After a careful scaling analysis, we determine that this is achieved by $\nu_{k}$ remaining constant in $L$, while $\sigma_F \sim 1/\sqrt{L}$. Under this scheme, we find $H$, $\varepsilon$, $P$, and other related quantities remain close to constant as $L$ increases.

\section{On the Interscale-flux and Dissipation Rate Distributions}

\begin{figure}
    \centering
    \includegraphics[width=8cm]{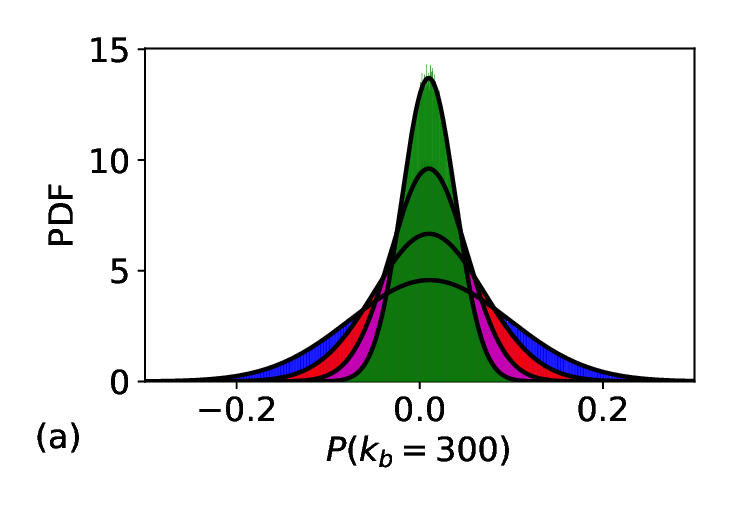}
    \includegraphics[width=8cm]{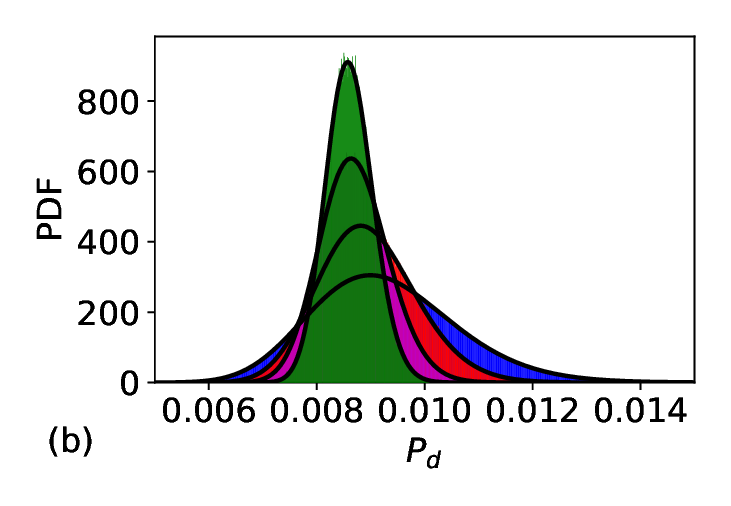}
    \caption{The distributions of $P(k_b,t)$ (a) and $P_d(t)$ (b) for the highest $\varepsilon$ data, for $L=2\pi$ (blue), $L=2\pi$ (red), $L=2\pi$ (magenta), and $L=2\pi$ (green). The distributions given by (\full) represent Gaussian (a) and Log-normal (b) distributions of equal mean and variance to the data they outline.}
    \label{fig:flux}
\end{figure}

In this section, we present our measurements of inter-scale energy flux, defined explicitly as $P(k_b = 300,t)$
\begin{equation}
    P(k_b ,t) = \sum_{|k|<k_b} \omega_k \frac{d|\hat{\psi}_{k}|^2}{dt},
    \label{eqn:flux_def}
\end{equation}
where the derivative on the RHS may be computed from the MMT equation given some statistically stationary field $\hat{\psi}_{k}(t)$. We begin with the flux distribution $P(k_b = 300,t)$, depicted in figure \ref{fig:flux}a of this supplemental material. Results only for the highest nonlinearity case are shown, however they are representative of the other cases. $P(t)$ takes a Gaussian distribution whose standard deviation is much larger than its mean. This is a similar result to our previous work in a 2D MMT model with a Nonlinear Schr{\"o}dinger Equation-like dispersion relation \cite{hrabski_properties_2022}, suggesting a Gaussian energy cascade is a general feature of wave turbulence. In fact, a simple argument supports a Gaussian energy cascade: first, consider the form of \eqref{eqn:flux_def}, which expresses the instantaneous inter-scale flux as a sum over a large number of modes. If the the instantaneous time rate of change of energy at each of these modes is assumed to be independent, then the Central Limit Theorem (CLT) can be used to show $P(k_b ,t)$ approaches a Gaussian distribution as the number of modes in the domain becomes large. Similarly, the inter-scale flux can be though of as a sum over a huge number of interactions, where again the CLT may be applied.

As $L$ is increased, the standard deviation of $P$ is decreased. Measurements of the standard deviation reveal that $\sigma(P(k_b ,t)) \sim 1/\sqrt{L}$. This too can be explained in the context of the CLT, as the variance of $P(k_b, t)$ is given by a large sum over $\sigma^2(\frac{d |\hat{\psi}_{k}|^2}{d t})$. $\sigma^2(\frac{d |\hat{\psi}_{k}|^2}{d t})$ scales linearly with the number of modes with $k < k_b$, which scales linearly with $L$. On the other hand, one can use the MMT equation to show that $\sigma^2(\frac{d |\hat{\psi}_{k}|^2}{d t}) \sim 1/L^2$, given that $|\hat{\psi}_{k}|\sim 1/\sqrt{L}$ by Parseval's theorem \eqref{eqn:parseval}. The end result is that $\sigma^2(P(k_b , t))\sim 1/L$, leading to the desired $\sigma(P(k_b, t)) \sim 1/\sqrt{L}$. 

In figure \ref{fig:flux}b, we provide the steady time distribution of high wave number dissipation rate, derived directly from the dissipation term to be 
\begin{equation}
    P_d (t) \equiv -\sum_{|k|>900}2\omega_k\nu_k|\hat{\psi}_k|^2.
    \label{eqn:pd_def}
\end{equation}
The distribution is log-normal, and the standard deviation exhibits the same $1/\sqrt{L}$ scaling due to similar arguments. Log-normal distributions of dissipation rate have been described in flow turbulence \cite{mandelbrot_intermittent_1974,pearson_log-normal_2018}, however we are not aware of similar results in WT.

\section{A Fast Direct Method for Computing Non-local Interactions}

To compute the non-local contribution to the small-scale energy evolution rate for a stationary spectrum, we begin by using a control volume argument to express the average dissipation rate directly in terms of the nonlinearity of the MMT model. This is achieved by
\begin{equation}
    P_d = J = -\sum_{|k| \ge 900}2\omega_k\nu_k \frac{dn_k}{dt} = -\sum_{|k| \ge 900}2\omega_k\nu_k\sum_{123} 2\text{Im}\langle \hat{\psi}_1 \hat{\psi}_2 \hat{\psi}_3^* \hat{\psi}_k^* \rangle \delta_K(k_1 + k_2 - k_3 - k).
    \label{eqn:Pd_nl}
\end{equation}
We note that for this section (in contrast to the previous section) $P_d$ indicates a time-averaged quantity. To proceed, we would like to compute the spectral evolution at some wavenumber $k\in \Lambda_L$ to interactions with the set of wave modes that reside in the forcing range. Let us call this set of forced modes $A\subseteq \Lambda_L$.  We will consider an interaction to involve $A$ if \emph{one or more} waves in the interaction are part of $A$. To express this clearly, it is useful to define the following set of 3-wave triples $(k_1,k_2,k_3)$:
\begin{equation}
    S^{(i)}_A \equiv \left\{(k_1,k_2,k_3)\ | \ (k_i \in A) \cup \left(\bigcup_{j\ne i}(k_j \in \Lambda_L)\right) \right\}.
    \label{eqn:Qi}
\end{equation}
$S_A^{(i)}$ denotes the set of all 3-wave triples where the $i$-th component is part of $A$ and the other two components are explicitly arbitrary. We are now ready to formally define $\left. \frac{d n(k)}{d t}\right|_A$, the spectral evolution involving interactions with $A$:
\begin{equation}
    \left. \frac{d n_k}{d t}\right|_A \equiv \sum_{(k_1,k_2,k_3)\in S_A^{(1)}\cup S_A^{(2)}\cup S_A^{(3)}} 2 \text{Im} \langle \hat{\psi}_1\hat{\psi}_2 \hat{\psi}^*_3 \hat{\psi}^*_k \rangle \delta_K (k_1 + k_2 - k_3 - k).
    \label{eqn:dnkdt_due_to_A}
\end{equation}
When substituted into \eqref{eqn:Pd_nl}, we will obtain the average dissipation rate due to interactions with the forcing range. While the definition of $\left.\frac{d n_k}{d t}\right|_A$ is clear in this form, it is expensive to compute. However, it is possible to rewrite the expression for the spectral evolution due interactions with $A$ as a carefully constructed series of sums which can be computed much more quickly.

We start by reformulating \eqref{eqn:dnkdt_due_to_A} in terms of the evolution of $\psi(k)$,
\begin{equation}
    \left. \frac{\partial n_k}{\partial t}\right|_A = \left\langle \psi_k \left.\frac{\partial \hat{\psi}^*_k }{\partial t}\right|_A + \hat{\psi}^*_k\left.\frac{\partial \hat{\psi}_k }{\partial t}\right|_A \right\rangle,
    \label{eqn:dnkdt2}
\end{equation}
with
\begin{equation}
    \left.i\frac{\partial \hat{\psi}_k }{\partial t}\right|_A \equiv \omega_k\psi_k +  \sum_{(k_1,k_2,k_3)\in S_A^{(1)}\cup S_A^{(2)}\cup S_A^{(3)}} \psi_{1}\psi_{2}\hat{\psi}^*_{3}\delta^{12}_{3k}.
    \label{dpsidt}  
\end{equation}
It follows from the inclusion-exclusion principle that the sum in the nonlinear term of (\ref{dpsidt}) can be rewritten as
\begin{align}
    \sum_{(k_1,k_2,k_3)\in S_A^{(1)}\cup S_A^{(2)}\cup S_A^{(3)}} = &\sum_{(k_1,k_2,k_3)\in S_A^{(1)}} + \sum_{(k_1,k_2,k_3)\in S_A^{(2)}} + \sum_{(k_1,k_2,k_3)\in
    S_A^{(3)}}    \label{inclusion-exclusion}  
 \\ - &\sum_{(k_1,k_2,k_3)\in S_A^{(1)}\cap S_A^{(2)}} - \sum_{(k_1,k_2,k_3)\in S_A^{(2)}\cap S_A^{(3)}} -
    \sum_{(k_1,k_2,k_3)\in S_A^{(1)}\cap S_A^{(3)}} \nonumber \\ + &\sum_{(k_1,k_2,k_3)\in S_A^{(1)}\cap S_A^{(2)}\cap S_A^{(3)}} \nonumber.
\end{align}
The advantage of this form is that sums over $S_A^{(i)}$ and intersections of $S_A^{(i)}$ can be computed quickly via Fourier transforms. Consider the term
\begin{equation}
    \hat{NL}(k,S_A^{(1)}) = \sum_{(k_1,k_2,k_3)\in S_A^{(1)}} \hat{\psi}_{1}\hat{\psi}_{2}\hat{\psi}^*_{3}\delta^{12}_{3k}.
    \label{eqn:NLK}
\end{equation}
We  already know that $\hat{NL}(k,\Lambda_L^3)$ (where the sum is taken over $(k_1,k_2,k_3)\in \Lambda_L^3$) is the Fourier domain representation of a quantity that is readily computed in physical domain, as this is exactly the basis of the pseudospectral method used to solve the MMT equation. Let's call the inverse Fourier transform of \eqref{eqn:NLK} $NL(x,S_A^{(1)})$. This can be written as
\begin{equation}
    NL(x,S_A^{(1)}) = \psi \psi^* \sum_{k_1\in A} \psi(k_1 )e^{ik_1  x}.
    \label{eqn:NLX}
\end{equation}
However, the last term in the product (\ref{eqn:NLX}) is simply the expression of an ideal band-pass filter applied to $\psi$ that admits only those modes in set $A$, and sets to $0$ any mode with $k\notin A$. Let $B_A$ be such a filter, so that
\begin{equation}
    NL(x,S_A^{(1)}) = \psi \psi^* B_A \left(\psi\right).
    \label{eqn:NLX2}
\end{equation}
By exactly the same argument, the sums over intersections may be computed. For example:
\begin{equation}
    NL(x,S_A^{(1)}\cap S_A^{(3)}) = \psi B_A \left(\psi^*\right) B_A \left(\psi\right).
    \label{eqn:NLX3}
\end{equation}
The inverse Fourier transform can then be used to obtain the desired term in the sum \eqref{inclusion-exclusion}. Every term in the sum \eqref{inclusion-exclusion} may be computed in physical domain according to this method, meaning (\ref{eqn:dnkdt2}) may also be computed in this way. For $N$ the number of modes, this method has $O(N\log N)$ complexity, as opposed to the $O(N^2)$ complexity of computing (\ref{eqn:dnkdt_due_to_A}) directly. The desired quantity $J_{nl} \equiv J_{A}$ for $A = [20,30]$ can then be computed, with
\begin{equation}
    J_A = -\sum_{|k| \ge 900}2\omega_k\nu_k \left.\frac{dn_k}{dt}\right|_A
    \label{eqn:Pd_nl2}
\end{equation}

Finally, we note that if one considers the set of interactions in $S_{A^C}^{(1)}\cap S_{A^C}^{(2)}\cap S_{A^C}^{(3)}$, where $A^C$ is the compliment to $A$, one can directly obtain
\begin{equation}
    \sum_{(k_1,k_2,k_3)\in S_A^{(1)}\cup S_A^{(2)}\cup S_A^{(3)}} = \sum_{(k_1,k_2,k_3)\in \Lambda_L^3} - \sum_{(k_1,k_2,k_3)\in S_{A^C}^{(1)}\cap S_{A^C}^{(2)}\cap S_{A^C}^{(3)}},
\end{equation}
where the RHS can also be evaluated in a pseudospectral way according to the method we have described.

\bibliography{HrabskiPanKinetic2023}